# Flow-Sensory Contact Electrification of Graphene

Xiaoyu Zhang[1], Eric Chia[1,†], Xiao Fan[1,†], Jinglei Ping[1,2,*]

[1]Department of Mechanical and Industrial Engineering, University of Massachusetts Amherst, Amherst, MA 01003, USA.

[2]Institute of Applied Life Sciences, University of Massachusetts Amherst, Amherst, MA 01003, USA.

* Corresponding author: Jinglei Ping

Address: 240 Thatcher Road LSL S603, Amherst, MA 01003

Phone: (413) 545-3395

Email: ping@engin.umass.edu

†These authors contributed equally to this work.

**Abstract**

All-electronic interrogation of biofluid flow velocity by sensors incorporated in ultra-low-power or self-sustained systems offers the promise of enabling multifarious emerging research and applications. Electrical sensors based on nanomaterials are of high spatiotemporal resolution and exceptional sensitivity to external flow stimulus and easily integrated and fabricated using scalable techniques. But existing nano-based electrical flow-sensing technologies remain lacking in precision and stability and are typically only applicable to simple aqueous solutions or liquid/gas dual-phase mixtures, making them unsuitable for monitoring low-flow (~micrometer/second) yet important characteristics of continuous biofluids (*e.g.*, hemorheological behaviors in microcirculation). Here we show that monolayer-graphene single microelectrodes harvesting charge from continuous aqueous flow provide an ideal flow sensing strategy: Our devices deliver over six months stability and sub-micrometer/second resolution in real-time quantification of whole-blood flows with multiscale amplitude-temporal characteristics in a microfluidic chip. The flow transduction is enabled by low-noise charge transfer at graphene/water interface in response to flow-sensitive rearrangement of the interfacial electrical double layer. Our results demonstrate the feasibility of using a graphene-based self-powered strategy for monitoring biofluid flow velocity with key performance metrics orders of magnitude higher than other electrical approaches.

**Main Text**

**Introduction**

Electrical transducers based on nanomaterials hold great potential for interrogating biofluid flow velocity in self-powered or ultra-low-power systems.[1-6] An electrokinetic approach[3-5] is to use nanotransistor devices to measure flow-dependent streaming potential/current. The devices are outstandingly miniaturized but subject to an intrinsic resolution limit of ~80 $\mu m\ s^{-1}$ (in a 1-Hz bandwidth) induced by thermal noise and liable to be reduced to ~20 $mm\ s^{-1}$ in a non-ideal measurement system. Alternatively, triboelectric charge harvested from a liquid flow by a micro/nanoelectrode device can be quantified for gauging the flow velocity. However, existing flowmeters based on this strategy typically use large-size (>$mm^3$) bundled nanotube/nanowire transducers[6] that are difficult to be scaled down, easy to cause flow-channel clogging, and prone to signal weakening and distortion due to the fouling of the electrodes upon specific/non-specific electrochemical processes and physicochemical adsorptions.[7, 8] Recently, flow sensors enabled by the cyclical formation of the electrical double layer (EDL) of the aqueous solution at a solid-aqueous interface have been developed but they are only suitable for liquid/gas dual-phase mixtures (cavity-confined solution,[9, 10] droplets,[11] and waving water[12]). Here we show self-powered graphene microdevices that transduce in real time the flow of continuous blood in a microfluidic channel to charge-transfer current in response to the flow-sensory rearrangement (not formation/deformation) of EDL at the graphene-aqueous interface. The devices deliver a resolution of $0.49 \pm 0.01$ μm s$^{-1}$ (in a 1-Hz bandwidth), a two-orders-of-magnitude improvement compared with existing device-based flow-sensing approaches (*Supplementary Information* Tab. S1), and are of low risk of being fouled or causing channel clogging. For periods exceeding six months the devices have demonstrated minimal variations in key performance metrics.

**Results and Discussion**

The flow transduction of the devices is based on a single microelectrode of monolayer graphene that harvests charge from flowing blood through contact electrification without the need for an external current supply. For implementing blood flow measurements, we fabricated acrylic chips with a graphene single-microelectrode device extending across over the microfluidic channel (Fig. 1**A** and *Supplementary Information* Fig. S1). The monolayer graphene was prepared via chemical vapor deposition (CVD) and transferred to the chip through a low-contamination electrolysis method.[13] The flow pathway, including the microfluidic chip and the tubes that connect to the chip, is entirely based on electrochemical inert materials. A flow of EDTA-anticoagulated whole bovine blood (pH

= 7.0, ionic strength = 150 mM) with precisely controlled velocity was driven through the microfluidic channel by a syringe pump. The graphene microelectrode was wired to the inverting input of an operational amplifier of a coulombmeter. The charge harvested from the solution by the graphene was stored in a feedback capacitor of the amplifier and quantified (Fig. 1**A**).

The charge transferred into a graphene device was measured as a function of time at various blood flow velocities. For each flow velocity, the transferred charge (Fig. 1**B**) indicates a clear, robust proportional relationship with respect to time. The charge-transfer current, obtained by proportional fit to the charge–time data, is of high precision (<10 fA) and signal-to-noise ratio (49.2 ± 0.7 dB), enabled by the low-noise characteristics of electron transfer at the interface of graphene with aqueous solution. The level of the current is minimally associated to the number of the graphene edge states (see *Supplementary Information* Fig. S2). According to the electron-transfer mechanism for solid/water contact electrification,[14] the charge transfer occurs via quantum tunneling of electrons through the electronic states of the defects (dangling bonds, electrical disorders, and grain boundaries) on the graphene basal plane,[15-17] entailing the ultra-low current noise level intrinsic to graphene-aqueous interfaces. Indeed, provided that the electrical conductance corresponding to a typical electron tunneling process is $\sim 10^{-24}$ $\Omega^{-1}$,[14] the specific conductance of the CVD graphene, $\sigma_{ct}$ (1.5 × $10^{-9}$ $\Omega^{-1}$ $mm^{-2}$),[18] leads to a defect density of $\sim 10^{17}$ $cm^{-2}$, in good agreement with that (1.44 × $10^{17}$ $cm^{-2}$) obtained from Raman spectroscopy.[18]

As shown by *Supplementary Information* Fig. S3, the charge-transfer current of the graphene device is linearly associated with the blood flow velocity, leading to a proportional relationship between the current response (the flow-induced current variation relative to the current at zero flow velocity) and the flow velocity (Fig. 2**A**, blue squares). The linear relationship covers over four orders of magnitude of flow velocity and is well reflected by the unit-slope power law of the current response normalized by the (global) sensitivity (the slope of the best proportional fit to the current–velocity data) as a function of flow velocity (Fig. 2**B**, blue squares).

Our control experiments indicate that the linear current–velocity relationship is not induced by piezoelectric effects (*Supplementary Information* Fig. S4) or charge transfer at the graphene edge electronic states (*Supplementary Information* Fig. S2). The effect of the streaming potential from the syringe to the graphene single microelectrode on the charge-transfer current at the graphene microelectrode is also negligible. First, the streaming potential depends on the length of the blood remained in the syringe during the measurement while in our experiment the measured current at the graphene microelectrode was independent of the length. Furthermore, the streaming potential is <2.7 µV according to $\Delta V \approx (\varepsilon \zeta \Delta p / \sigma \eta)$, where $\varepsilon$ is the blood permittivity (7.68 × 8.854 × $10^{-12}$ F m$^{-1}$), $\zeta$ the zeta potential (<60 mV in magnitude), $\Delta p$ the pressure drop (1650 Pa), $\sigma$ the blood electrical conductivity (~1.23 S m$^{-1}$), and $\eta$ the blood viscosity (~3.5 cP).[19] Adding this potential to the electrical potential at the graphene EDL generates negligible, <2.0-fA charge-transfer current at the graphene (provided that the charge-transfer resistance, $R_{ct}$, is ~ 1.3 GΩ).[18] Considering the electron-tunneling origin of the contact-electrification current, the flow transduction can be understood by the flow-induced rearrangement of the EDL at the graphene basal-plane defects: An increase in the blood flow velocity increases the wall shear stress generated by the flow at the graphene-aqueous interface, suppresses the counterions' screening effect,[20] and reduces the electrical potential barrier at the blood side of the graphene-defect electronic states, resulting in enhancement in the transferring of electrons from the graphene to the blood, as shown by the energy profile in Fig. 1**B**. As the blood flow increases in velocity, this effect reverses the polarity of the current and then increases its magnitude (*Supplementary Information* Fig S3).[21] To the first order of magnitude, the impact of flow velocity $v$ on the charge-transfer current $i$ can be written as $i \approx i_0 + sv$, where $i_0$ is the baseline current (the current at zero flow velocity) and $s$ the sensitivity.

The first-order EDL-rearrangement scenario for the flow-sensory charge-transfer current of the graphene devices suggests that the sensitivity of the device can be improved by modulating the electrical potential distribution of the EDL at the graphene/blood interface to enhance the charge-transfer current thereat. Biased non-faradaic counter electrodes made of inert metals, such as platinum, are typically unstable and can generate substantial (>10–100 pA) background noise to the flow-induced charge-transfer current signal measured by the graphene microelectrode, thereby burying the signal in the noise. In order to control the charge-transfer current at an accuracy of sub-pA level, we used an unbiased non-faradaic electrode made of glass with variable surface charge density that can generate correspondingly different ultra-stable electrical potential coupling to the graphene EDL potential structure. As shown in the energy profile diagram in Fig. 2**C**, at $v = 0$, the adsorption of specific counterions renders the electrostatic potential reversed at the glass EDL[22-24] and reduces the electrical potential barriers on both the blood side and the graphene side at a graphene disorder electronic state to different extents. The result is that the potential barrier on the graphene side is relatively higher than that on the blood side and more electrons transfer out from the graphene to the blood compared with the situation without the modulation. In comparison with new glass, the surface charge density of aged glass is higher, leading to greater change in the energy profile and, in consequence, more enhanced baseline charge-transfer current of the graphene devices (Fig. 2**C**).

Figure 2**A** and **B** and *Supplementary Information* Fig. S5 show that at each modulating level represented by a corresponding baseline current, the current response of the graphene device is proportional to the flow velocity, as in the unmodulated experiment, aside from minimal deviations at flow velocities <0.1 mm s$^{-1}$ induced by the modulation. An enhanced charging status (higher baseline current magnitude) of the glass electrode corresponds to higher magnitude of the charge-transfer current response of the graphene device and higher magnitude of the slope of the current-velocity data, *i.e.*, higher sensitivity for flow velocity determination (Fig. 2**D**). The sensitivity enhancement delivers an optimal resolution (in a 1-Hz bandwidth) of 0.49 ± 0.01 μm s$^{-1}$ (Fig. 2**D** and *Supplementary Information* Fig. S6), a two-orders-of-magnitude improvement compared with previous device-based flow sensors (*Supplementary Information* Tab. S1). The remarkable resolution of the flow-velocity measurement is imparted by the low-noise electron transfer of graphene in contact with the blood flow. The log-log plot of the flow-

velocity resolution vs. bandwidth relationship of our measurement (Fig. 2**E**) shows a clear power law. The slope of the best linear fit is 0.47±0.01, suggesting a square root dependency of the current noise with bandwidth, in good agreement with the thermal noise arising from the charge-transfer resistance at the graphene-water interface: $i = \sqrt{4k_B T \Delta f / R_{ct}} = 3.56\sqrt{\Delta f}$ (fA Hz$^{-1/2}$), where $k_B$ is the product of the Boltzmann constant, $T$ the temperature (298 K), and $\Delta f$ the bandwidth.

Under the first-order EDL-rearrangement scenario, the graphene charge-transfer current modulation can be modeled by adding a dependent current source (with a scaling factor β) to the linear circuit description of the graphene/blood interface (Fig. 2**A**): $i \rightarrow i + \beta i$. Considering the current–flow relationship ($i = i_0 + sv$), $i_0$ and $s$ are multiplied by a same factor: $i_0 \rightarrow (1 + \beta)i_0$ and $s \rightarrow (1 + \beta)s$, yielding a proportional $i_0$–$s$ relationship over the whole range of $i_0$, in good agreement with Fig. 2**D**. The slope (0.130 ± 0.004 mm s$^{-1}$) of the best proportional fit represents a common blood flow velocity corresponding to zero charge-transfer current for all modulation levels. This property is useful for device calibration when the modulation is at an unknown level.

We also measured the contact-electrification charge transfer for the flow of phosphate buffered saline (PBS) at different velocities by using the same graphene device that was used in the blood flow measurement. The transferred charge shows clear linearity with respect to time for each flow velocity and the current response is proportional to the flow velocity (*Supplementary Information* Fig. S7), as in the blood-based experiment. The magnitude of the sensitivity of the device for PBS velocity determination is 4.6× smaller than that of whole blood due to the lower viscosity of PBS that generates lower wall shear stress impacting on the EDL rearrangement at the graphene/PBS interface.

We then investigated the repeatability and long-term stability of the graphene devices and their capability for real-time identification of multiscale amplitude-temporal flow characteristics. We developed a program for automatically picking up charge-transfer signal from a graphene device and providing the corresponding flow-velocity readout. The program (*Supplementary Information* Video S1) functions in real time to communicate with the coulometer to acquire charge-transfer data, extract electrical current by taking numerical time derivative of the transferred charge, smooth the current using a bandwidth-controllable Savitzky-Golay filter, and translate the smoothed current to flow velocity via linear interpolation of the current–flow data of the graphene device being used.

Figure 3**A** shows the real-time flow velocity measured by a graphene device in response to a continuous five-step blood flow that lasted for more than two hours. The measured velocity demonstrates high repeatability with minimal fluctuations ($\pm$ 0.07 mm s$^{-1}$). We also used a graphene device to perform intermittent measurements (cleaned with PBS after each measurement) for periods of six months. The blood-flow sensitivity of the device fluctuated around an average value (0.39 pA s mm$^{-1}$) with a standard deviation of $\pm$ 0.02 pA s mm$^{-1}$, $\pm$ 5.1% of the average value, as shown in Fig. 3**B**. The repeatability, stability, and anti-fouling capability of the graphene devices reflected by the experiments are outstanding, despite that graphene is only a monolayer layer of carbon atoms.

Intrinsically, graphene is of atomic thickness and flatness, making it an ideal material to be fabricated into the planar device configuration easily to be integrated in the microfluidic device with minimal risk of being fouled or clogging the microfluidic channel. Graphene is formed by C–C sp$^2$-hybridized σ bonds with π-band electrons oriented out of the basal plane, which give rise to high physicochemical stability and high electrochemical inertness in the solutions with physiological conditions (pH 7.0, ionic strength 150 mM, room temperature, *etc.*).

We used the graphene devices to register rapid, multiscale variations in four pulsatile waveforms of blood flow driven by the syringe pump, the sawtooth-like (0.33–6.00 mm s$^{-1}$; Fig. 4**A**), the sinuous (1.60–1.66 mm s$^{-1}$; Fig. 4**B**), the human-retinal-capillary (1.0–2.0 mm s$^{-1}$; Fig. 4**C**),[25] and the murine-brain-capillary (9.0–23.0 μm s$^{-1}$; Fig. 4**D**),[26] in addition to a pulsatile waveform of PBS flow (*Supplementary Information* Fig. S8). For all the waveforms, the real-time readout of our program well follows the set flow velocity with minimal delay. In comparison, the outcome of a state-of-the-art high-sensitivity calorimetric flowmeter can deviate substantially from the set flow velocity (Fig. 4**B**), despite that calorimetric flowmeters have been broadly used in low-flow microfluidic research and applications. For the waveform that simulates the blood in murine deep-brain capillaries (Fig. 3**D**), the standard deviation of the readout with respect to the set flow velocity is 1.2 μm s$^{-1}$, representing a notable performance metric of the graphene devices in identifying small-scale flow features over other electrical flow-sensing technologies. This flow resolution is comparable with high-performance imaging modalities such as real-time fluorescent cerebrovascular imaging (5–20 μm s$^{-1}$)[26] and high-speed adaptive optics imaging (18–70 μm s$^{-1}$),[25, 27] while the graphene devices retain further advantages in the long-term stability, miniaturization, integrability, and cost efficiency.

Our research provides a self-powered strategy for high-performance biofluid-flow interrogation enabled by graphene. The findings pave the way to future researches on all-electronic *in vivo* flow monitoring in investigating new, ultra-low-flow (< 10 μm s$^{-1}$) life phenomena in metabolomics, retinal hemorheology, and neuroscience. Potential fields of applications of our technology include *in vivo* biofluid mechanics, vascular tissue engineering, and disease-progression surveillance, to name a few,[1, 2, 28-30] particularly when established imaging modalities are less efficient or difficult to be implemented. The charges transferred into the graphene devices in the measurement are stored in a feed-back capacitor, offering a pathway to highly self-sustained systems that simultaneously probe blood flow and harvest energy from the flow.

**Materials and Methods**

**Graphene Preparation and Transferring**

We prepared monolayer graphene *via* CVD on copper catalytic substrate. A piece of copper foil (99.8% purity; Alfa Aesar) was loaded into a quartz tube (22 mm in I.D.; 4 feet in length) and annealed for 30 min at 1060 °C in the atmosphere of mixed ultrahigh purity (UHP; 99.999%) hydrogen (flow rate = 200 sccm) and UHP (99.999%) argon (flow rate = 500 sccm) for removing oxide residues. Then ambient-pressure chemical vapor deposition was implemented for the growth of monolayer graphene in the mixed gas of UHP hydrogen (flow rate = 3 sccm), UHP argon (flow rate = 500 sccm), and precursor UHP (99.99%) methane (flow rate = 0.5 sccm) at 1035 °C (growth time 20 min).

A piece of the CVD graphene (100–1000 μm in width) was transferred from the copper substrate onto a 0.5-mm thick acrylic sheet using the low-contamination bubbling method. At first, a 400-nm thick layer of polymethyl methacrylate (MICROCHEM, 950 PMMA A4) was spun-coated on the CVD graphene on copper substrate at a speed of 1000 rpm. Then the PMMA/graphene/copper film was connected to the cathode of a power supply and immersed in 1M NaOH solution. An electric current of ~1 A was applied through the film to generate hydrogen bubbles between the graphene and copper and float off the PMMA/graphene film from the copper. The PMMA/graphene film was then transferred onto a 0.5 mm thick acrylic sheet with the PMMA in contact with the acrylic sheet and the graphene facing out, forming a graphene/PMMA/acrylic structure.

**Device Fabrication**

A GCC LaserPro Spirit GLS laser engraver was used to fabricate 500 × 500 μm microchannels in an acrylic sheet (500 μm in thickness). Then the microchannel-structured acrylic sheet and a 0.5-mm thick substrate acrylic sheet were thermally bonded with the assistance of a solution of 80 vol% dichloromethane (DCM, Honeywell Burdick &

Jackson 300-4) and 20 vol% 2-propanol (IPA, Fisher Chemical A416-4) at 70 °C for 10 minutes in a forced convection oven, forming a structured microfluidic module.

We used e-beam evaporation to deposit a 60-nm thick Cr/Au lead through which the transferred graphene on the graphene/PMMA/acrylic structure can be electrically connected with a coulombmeter. Then the microfluidic module and the graphene/PMMA/acrylic structure were thermally bonded with the assistance of the DCM/IPA solution for 90 seconds at about 100 °C using a heat gun. In this process, the PMMA film sandwiched between the graphene and the acrylic module was bonded to the acrylic module. The device was then cleaned by DI water in an ultrasonic bath for 10 minutes for residual removal.

**Coulometric Measurement**

In measurement, the graphene single microelectrode in the microfluidic chip was exposed to 1×PBS (Fisher BP661-50; pH = 7.0, ionic strength = 150 mM) or bovine whole blood (BIOIVT BOV7411) with flow velocity controlled by a picoliter/min syringe pump (Harvard Apparatus 70-3009). The bovine blood was stored at -86 °C and thawed in a water bath to room temperature before the measurement. Only the blood samples underwent one freeze–thaw cycle were used. The PBS solution was prepared before the experiment.

A Keithley 6517b electrometer in the coulombmeter mode (1-fC charge-measurement resolution, settling time <0.1 μs) was used to quantify the charge transfer of the graphene single microelectrode. Prior to the measurement, the noninverting input of the coulombmeter was grounded while the inverting input was connected to a dissipative resistor. To start the measurement, the inverting input was disconnected from the dissipative resistor and connected to the Au/Cr lead, which linked the graphene single microelectrode to a virtual ground so that the charges that transferred into the graphene were stored in a feed-back capacitor of the operational amplifier in the coulombmeter and quantified.

We developed a program to acquire the flow-sensory charge signal from the coulometer through an IEEE-488 interface. The program performs time-derivative of the measured charge to obtain the charge-transfer current and uses a bandwidth-variable Savitzky-Golay filter based on the least-squares polynomial method to smooth the current in real time. The measured electric current is translated to flow velocities in the program in real time by interpolating the current–flow data set of the device being tested. The flow velocity measured by our graphene devices was in real time compared to that measured by a state-of-the-art high-sensitivity calorimetric flowmeter (Sensirion SLI1000).

**Author Contributions**

J.P. and X.Z. conceived and designed the project. E.C. and X.Z. fabricated the devices. X.F. prepared graphene samples. X.Z. conducted the measurement. J.P. and X. Z. analyzed the experimental results and wrote the manuscript. All authors discussed the results and commented on the manuscript.

**Acknowledgments**

The authors acknowledge J. Nicholson and J. Yin for e-beam evaporation in UMass Amherst Silvio O. Conte National Center. J.P. acknowledges support from Department of Defense (DoD), Air Force Office of Scientific Research (AFOSR) (FA9550-20-1-0125); DoD, Congressionally Directed Medical Research Programs (CDMRP) (W81XWH-19-1006).

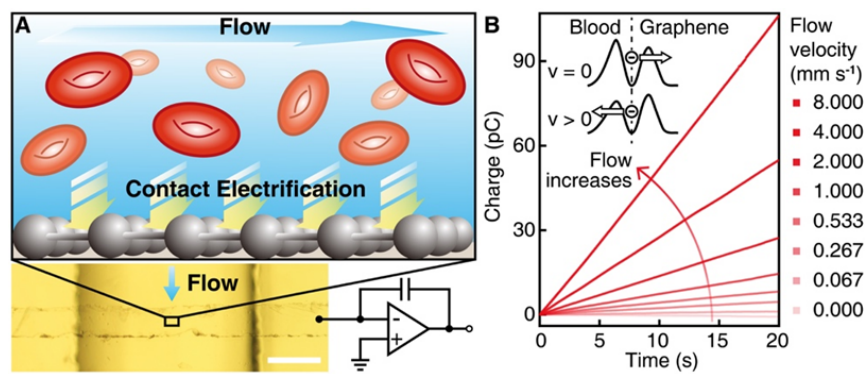

**Figure 1. Transducing blood flow velocity to electric current by using a graphene single-microelectrode device. A.** Coulometric measurement of contact-electrification charge transfer between whole blood and graphene. The optical image shows a monolayer-graphene microelectrode crossing over the microfluidic channel in an acrylic chip. The scale bar is 200 μm. **B.** The measured unsmoothed charge transfer of a graphene device as a function of time for different blood-flow velocities. The diagram illustrates the electron transfer through the electronic state of a graphene defect site. The arrows represent the directions of the net electron flow.

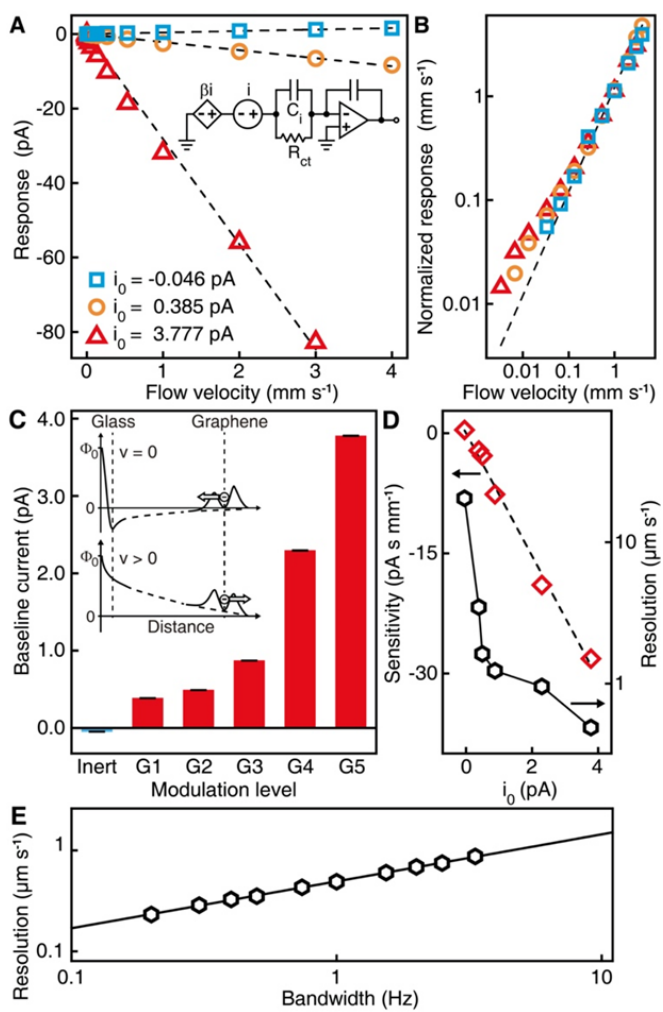

**Figure 2. Response curves and characteristics for blood flow-velocity quantification by the graphene single-microelectrode device. A.** The current response as a function of flow velocity. The dash lines are best proportional fits to the data. The linear electrical circuit models the charge-transfer current through the graphene/blood interface represented by a charge-transfer resistance ($R_{ct} = 1/\sigma_{ct}$) and an interfacial capacitance ($C_i$). **B.** The sensitivity-normalized current response as a function of flow velocity. The eye-guiding dash line is of unit slope. The data symbols are the same as in **A**. **C.** The baseline current (at zero flow velocity) measured with (G1–5) and without (Inert) using an unbiased glass electrode. The energy profile diagram illustrates the modulating effect

of the charged glass electrodes on the_graphene–blood electron transfer through a graphene-defect electronic state. **D.** Resolution and sensitivity as a function of the baseline current $i_0$. The dash line is a proportional fit to the sensitivity data. **E.** Resolution as a function of bandwidth for $i_0$ equal to 3.777 pA (G5). The black line is based on the parameters of the best linear fit to the data. In **A**, **B**, **D**, and **E**, the sizes of the error bars are smaller than the size of the data points.

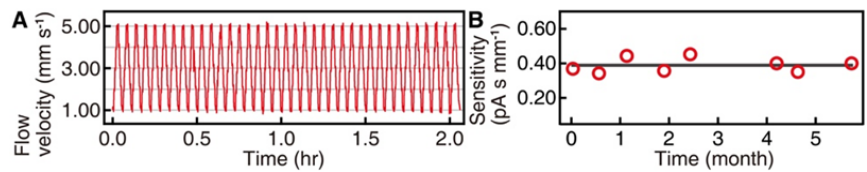

**Figure 3. Repeatability and stability of the graphene device. A.** The measured flow velocity in response to a stepwise flow waveform switching between 1, 2, 3, 4, and 5 mm s$^{-1}$ every 20 seconds in turn. **B.** Long-term (half-year) stability of sensitivity. The sizes of the error bars are smaller than the size of the data points. The black line is a constant fit to the data.

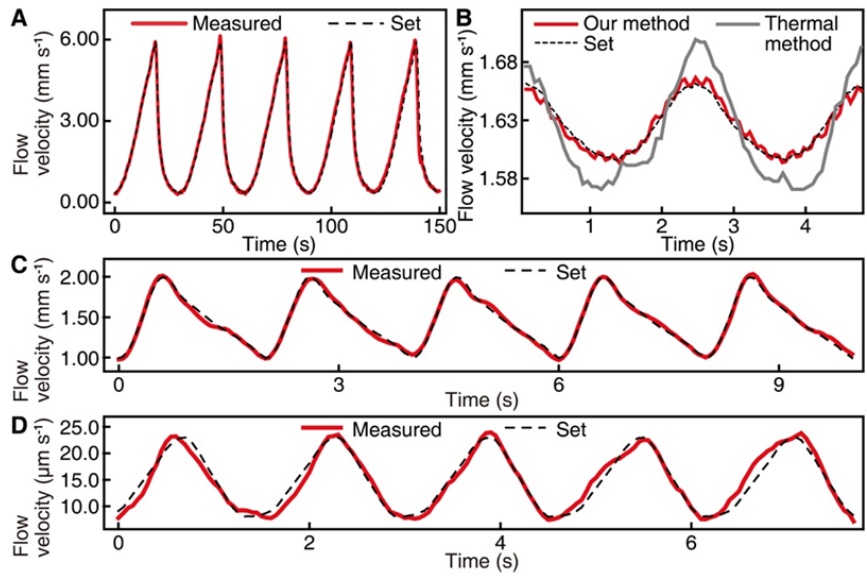

**Figure 4. Real-time interrogation of pump-driven time-varying whole blood flows using the graphene devices.** The measured velocities of the sawtooth-like (**A**) and sinuous (**B**) flows and the flows whose waveforms simulate those of the blood flows through murine deep-brain capillaries (**C**) and human retina capillaries (**D**). The velocity of the sinuous flow (**B**) measured by the graphene device shows 6-Hz, <10 μm s$^{-1}$ periodic steps generated by the stepping motion of the stepper motor of the syringe pump.

# Flow-Sensory Contact Electrification of Graphene

Xiaoyu Zhang[1], Eric Chia[1,†], Xiao Fan[1,†], Jinglei Ping[1,2,*]

**Affiliations**

[1]Department of Mechanical and Industrial Engineering, University of Massachusetts Amherst, Amherst, MA 01003, USA.

[2]Institute of Applied Life Sciences, University of Massachusetts Amherst, Amherst, MA 01003, USA.

[†]These authors contributed equally to this work.

*Corresponding author: ping@engin.umass.edu.

Table of Contents

**Table S1. Summary of typical device-based flow sensors developed in previous researches**.

| Authors | Resolution/ Detection Limit (mm s$^{-1}$) | Transduction Mechanism | Tested Fluid |
|---|---|---|---|
| Bertrand Bourlon, Marc Bockrath, et al.[1] | 20.8 | Streaming potential | NaCl solution |
| Rong Xiang He, Feng Yan, et al.[2] | 4.62 | Streaming potential | KCl solution |
| A. K. M. Newaz, K. I. Bolotin, et al.[3] | 0.1 | Streaming potential | NaCl solution |
| Dong Rip Kim, Xiaolin Zheng, et al.[4] | 0.55-1.1 | Streaming potential | KCl solution |
| Ying Chen, J. Iwan D. Alexander, et al.[5] | ~3.0 | Streaming potential | DI water |
| Shota Sando, Tianhong Cui, et al.[6] | 0.59 | Streaming potential | PBS |
| B. H. Son, Y. H. Ahn, et al.[7] | 0.42 | Streaming potential | DI water |
| Xiuhan Li, Zhong Lin Wang, et al.[8],* | ~0.2 | Triboelectricity | Tap and DI water |
| Jun Yin, Wanlin Guo, et al.[9],* | 13 | Triboelectricity | $NH_3 \cdot H_2O$, NaCl, $MgCl_2$, HCl solution, water |
| Jun Yin, Wanlin Guo, et al.[10],* | 22.5 | Triboelectricity | LiCl, NaCl, KCl, HCl, NaF, NaBr |
| Trevor Hudson, Ellis Meng, et al.[11] | 0.035 | Calorimetry | PBS |
| Jaione Etxebarria, Aitor Ezkerra, et al.[12] | 0.013 | Calorimetry | DI water |
| Alex Baldwin, Ellis Meng, et al.[13] | 0.019 | Calorimetry | PBS |
| Ellis Meng, Yu-Chong Tai, et al.[14] | 0.017 | Calorimetry | DI water |
| R Ahrens, K Schlote-Holubek, et al.[15] | 0.625 | Calorimetry | Water, oil |
| Lingling Zhang, Xingzhong Zhao, et al.[16] | 0.06 | Piezoelectric effect | PBS |

* Those flow-sensory approaches are not suitable for continuous flows.

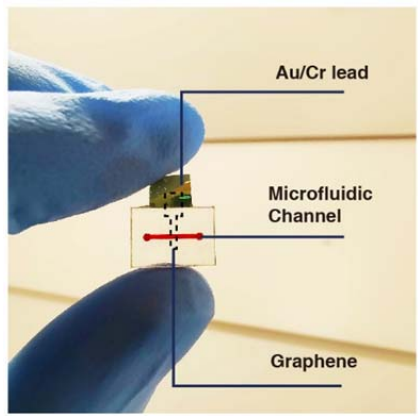

**Figure S1. An acrylic-based microfluidic chip that incorporates a graphene single microelectrode.** The Au/Cr lead was e-beam deposited for electrically connecting the graphene microelectrode with a coulombmeter for charge measurement. Liquid was infused by a syringe pump through the microchannel that crosses over the graphene microelectrode.

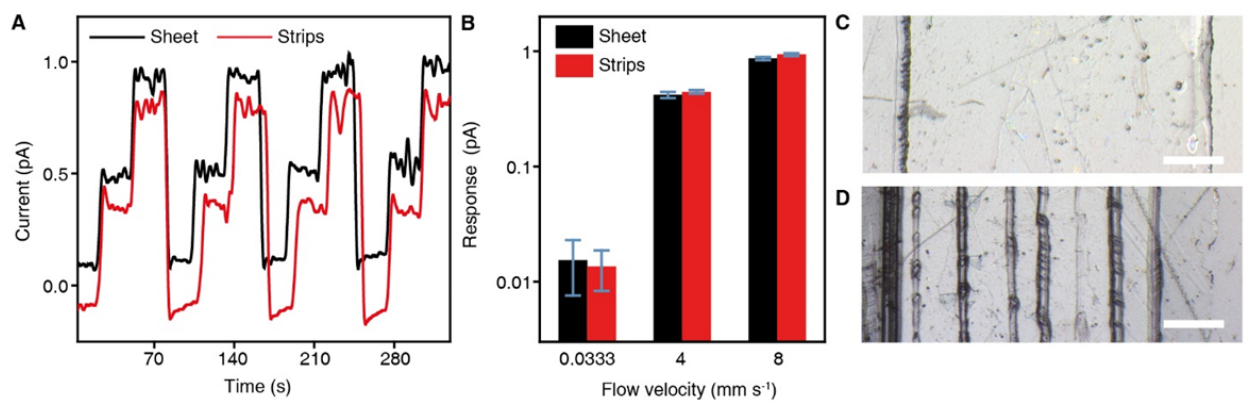

**Figure S2. The charge transfer of graphene is minimally associated to the graphene edge electronic states. A.** The charge-transfer current that was measured by using a 1-mm wide sheet of graphene (with two edges) and that by using graphene stripes (with 16 edges) that were made from the graphene sheet by precision cutting. The syringe pump speed was changed in turn between 0.033, 4, and 8 mm s$^{-1}$. **B.** The creation of the edges neither increases the calibration current magnitude nor the derivative of the current response with respect to flow velocity (the sensitivity). **C.** The optical image of the graphene sheet prior to being cut. **D.** The optical image of the graphene strips that were precision cut from the graphene sheet. The scale bars in **C** and **D** are 200 µm. The results suggest that the charge-transfer current and the current–flow association arise from the graphene basal plane electronic states.[17-19]

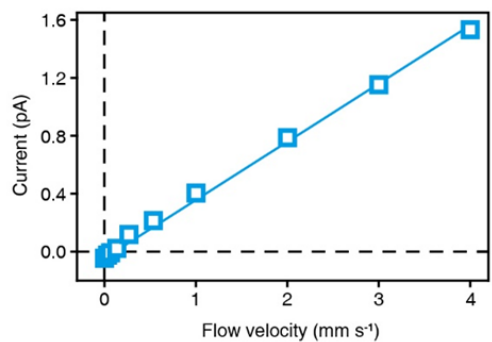

**Figure S3. The charge-transfer current as a function of flow velocity. The current was obtained by proportional fit to the charge–time data (Fig. 1B).** The solid line is the best linear fit to the data. The intersection of the eye-guiding dash lines is the origin of the coordinate system. The sizes of the error bars are smaller than the size of the data points.

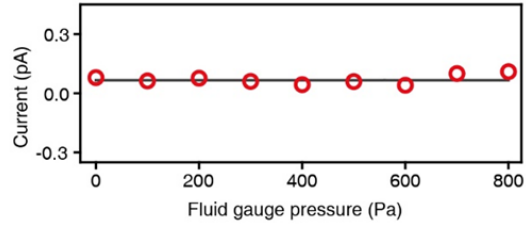

**Figure S4. Charge-transfer current as a function of pressure difference for stagnant blood over a graphene device.** The blood was kept motionless for each applied gauge pressure in the measurement. The black line is the best constant fit to the data. The fit value is $0.067 \pm 0.008$ pA. The sizes of the error bars are smaller than the size of the data points. Since the maximum gauge pressure difference over the graphene device in our flow-varying charge transfer measurements is $665 \pm 44$ Pa (corresponding to a flow velocity of 8 mm s$^{-1}$), the flow-response current is minimally associated with potential piezoelectricity effects.[20]

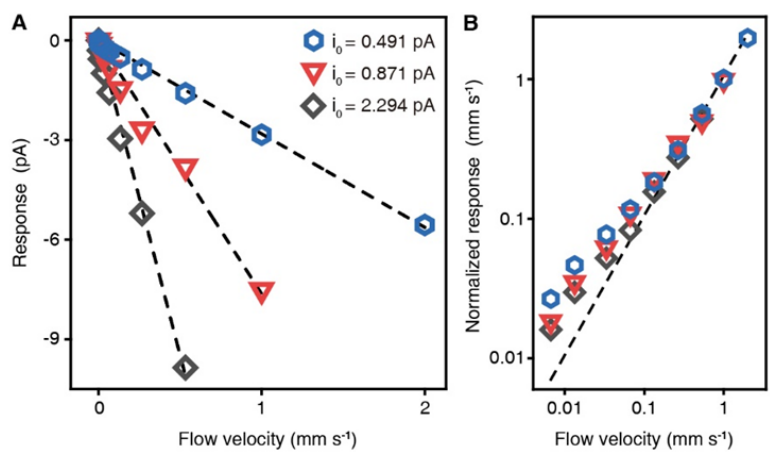

**Figure S5. Current response of the graphene device in addition to Fig. 2A and B. A.** The current response as a function of flow velocity. The dash lines are best proportional fits to the data. **B.** The sensitivity-normalized current response as a function of flow velocity. The eye-guiding dash line is of unit slope. The data symbols are the same as in **A**. In **A** and **B**, the sizes of the error bars are smaller than the size of the data points.

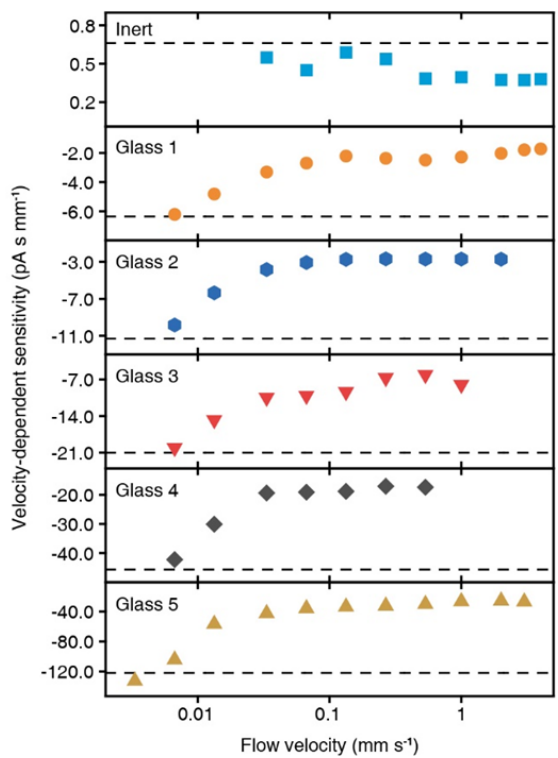

**Figure S6. The velocity-dependent sensitivity for measurements with (Glass1–5) and without (Inert) using an unbiased glass electrode.** The velocity-dependent sensitivity is obtained by taking numerical derivative of the current with respect to flow velocity at different flow velocities. The dash lines are the baseline velocity-dependent sensitivities (at zero flow velocity). The optimal resolution is obtained as the ratio of the standard deviation of the measured current to the absolute value of the baseline velocity-dependent sensitivity. The descriptions for the glass electrodes are the same as in *Supplementary Information* Fig. S5B. The sizes of the error bars are smaller than the size of the data points for all panels.

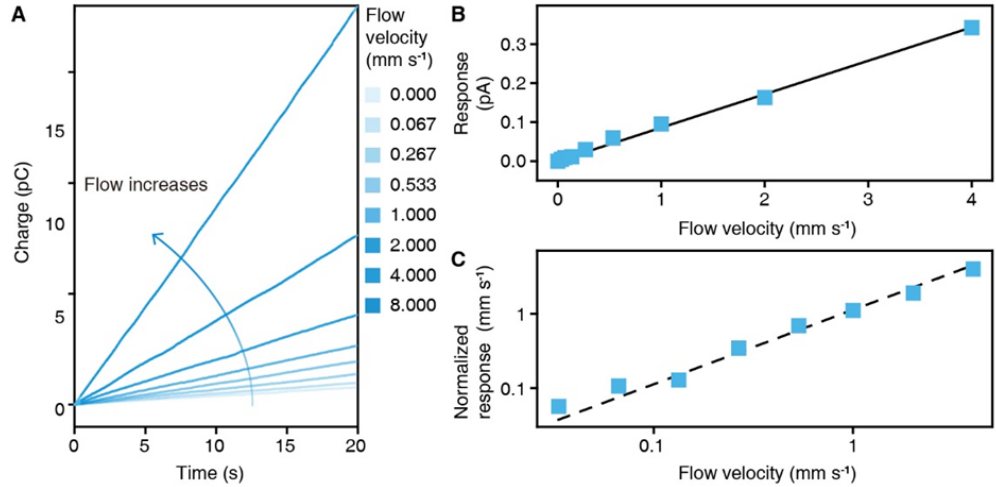

**Figure S7.** Real-time charge transfer, current response, and the normalized current between the graphene single microelectrode and PBS in measurement without using an unbiased glass electrode. **A.** Real-time unsmoothed charge transfer. The graphene device is the same one that was used in the blood measurement. **B.** The relationship between the current response and flow velocity. The black line is the best proportional fit to the data, which yields the fit parameter value of the slope (the sensitivity of the device) equal to $0.086 \pm 0.002$ pA s mm$^{-1}$, $0.22 \pm 0.01$ times of that for whole blood. The currents were extracted in a bandwidth of 1 Hz. **C.** Log-log plot for the normalized current versus flow velocity. The eye-guiding dash line is of unit slope. The sizes of the error bars in **B** and **C** are smaller than the size of the data points.

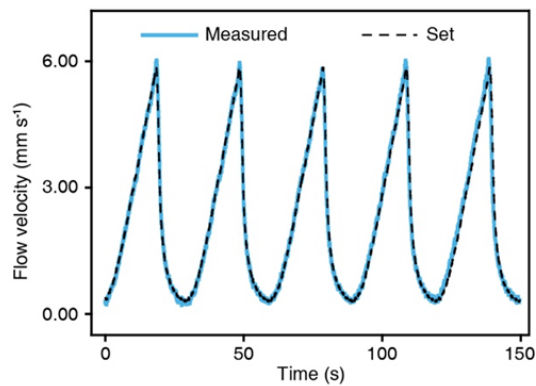

**Figure S8. Monitored flow velocity in response to sawtooth-like PBS flow waveform driven by the syringe pump.** The measured velocity in *A* and *B* responds with minimal delay to abrupt variations at the sharp apexes in the waveforms and the discrepancy of the flow velocity with the set velocity is minimal.

**Video S1. The program for real-time signal acquisition and data processing.**

The panel at the top right corner shows the electrical current that is numerically extracted from the transferred charge (lower right corner) of a graphene single-microelectrode device in real time. The current is smoothed by a real-time digital S-G filter (in a 1-Hz bandwidth). The left panel shows the flow velocity that is translated from the smoothed current by implementing interpolation of the corresponding current–flow data set.